\documentclass{PoS}

\title{Annihilation vs. Decay: Constraining Dark Matter Properties
  from a Gamma-Ray Detection in Dwarf Galaxies} 

\ShortTitle{Annihilation vs. Decay}

\author{\speaker{Sergio Palomares-Ruiz} \thanks{This work was partially
    supported by the Portuguese FCT through CERN/FP/109305/2009 and
    CFTP-FCT UNIT 777, which are partially funded through POCTI
    (FEDER), and by the Spanish Grant FPA2008-02878 of the MICINN.}\\
        Centro de F\'{\i}sica Te\'orica de Part\'{\i}culas (CFTP),\\
        Instituto Superior T\'ecnico, \\ 
        Avenida Rovisco Pais 1, 1049-001 Lisboa, Portugal \\
        E-mail: \email{sergio.palomares.ruiz@ist.utl.pt}}

\author{Jennifer M. Siegal-Gaskins\\
        Center for Cosmology and Astro-Particle Physics (CCAPP),\\
        The Ohio State University, \\
        191 W. Woodruff Ave., Columbus, OH 43210 USA \\
        E-mail: \email{jsg@mps.ohio-state.edu}}

\abstract{\begin{flushright}
CFTP/10-017\\
\end{flushright}

Although most proposed dark matter candidates are stable, in order for
dark matter to be present today, the only requirement is that its
lifetime is longer than the age of the Universe, $t_U \simeq 4 \times
10^{17}$~s.  Moreover, the dark matter particle could be produced via
non-thermal processes and have a larger annihilation cross section
from the canonical value for thermal dark matter, $\langle\sigma
v\rangle \sim 3 \times 10^{-26}$~cm$^3$~s$^{-1}$.  We propose a strategy 
to distinguish between dark matter annihilation and/or decay in the
case that a clear signal is detected in future gamma-ray observations
of Milky Way dwarf galaxies with gamma-ray experiments.  The
discrimination between these cases would not be possible in the case
of the measurement of only the energy spectrum.  We show that by
studying the dependence of the intensity and energy spectrum on the 
angular distribution of the signal, the origin of the signal could be
identified, and some information about the presence of substructure
might be extracted. 
}

\FullConference{Identification of Dark Matter 2010-IDM2010\\
		July 26-30, 2010\\
		Montpellier France}

\begin{document}

\section{Introduction}
\label{introduction}

The existence of dark matter (DM) is inferred from many different
astrophysical and cosmological observations, which indicate that it
constitutes about 80\% of the mass content of the Universe.  However,
aside from its gravitational interactions, very little is known about
its nature.  Among the many proposed particle candidates,  the most
popular one are weakly interacting massive particles (WIMPs) with
masses in the range 10~GeV--10~TeV. Although most proposed WIMPs are
stable and are produced thermally in the early Universe with an
annihilation cross section (times relative velocity) of $\langle\sigma
v\rangle \sim 3 \times 10^{-26}$~cm$^3$~s$^{-1}$, DM may be unstable
but long-lived, with a lifetime $\tau_\chi$ much longer than the age
of the Universe $t_{\rm U} \simeq 4 \times 10^{17}$~s.  Moreover, DM
might have been produced via non-thermal processes and have a larger
annihilation cross section than the canonical value for WIMP thermal
relics.

Among the different ways to detect DM, indirect searches look for the
products of DM annihilation or decay, which include antimatter,
neutrinos and photons.  During the last years, different approaches 
have been proposed to constrain dark matter properties by 
using indirect measurements~\cite{indirectnuprop, indirectgamma,
  Hensley:2009gh}. However, to extract the properties of the DM
particle from the detection of an indirect signal requires several
pieces of information. There exist many different degeneracies among
the different parameters which determine the energy spectrum of the
signal. In general, this prevents accurate reconstruction of the DM
properties from the energy spectrum alone. In particular, the sole
measurement of the energy spectrum would make it impossible to know if
the indirect signal from DM is produced by annihilation or decay.  The
spectrum of the former is characterized by a cutoff at an energy equal
to the DM mass, while the cutoff in the spectrum from the latter is at
an energy equal to half of the DM mass.

In this talk (see also Ref.~\cite{PalomaresRuiz:2010pn}), we address
the question whether annihilation and/or decay can be identified as
the origin of a DM signal in gamma rays.  We note that if DM is
unstable and produces an observable signal from decay, an annihilation
signal will also be present.  We show that there is a range of
parameters for which the two signals would be comparable, and in this
case, angular information could help to determine their presence and
their relative contribution to the total signal. Although very
challenging, this would identify DM as an unstable particle.

In particular, in order to tackle this problem, we suggest a strategy
to distinguish between these scenarios using future gamma-ray
observations of Milky Way dwarf galaxies.  We show that, in the case
that a gamma-ray signal is clearly detected, the origin could be
identified as DM decay, annihilation, or both by examining the
dependence of the intensity and energy spectrum on the angular
distribution of the emission.  Furthermore, if annihilation and decay
each contribute significantly to the signal, we show how these
observations could be used to extract information about the DM mass,
lifetime, annihilation cross section, and dominant annihilation and
decay channels.  In addition, as a byproduct of this analysis, one
might also establish or limit the contribution to the signal from
substructure in the dwarf galaxy's halo.

\section{General Idea}

An indirect signal from annihilation or decay originates from the same
DM particles, but these two processes give rise to different angular
distributions of the emission and different energy spectra.  As
pointed out in Refs.~\cite{Bertone:2007aw, PalomaresRuiz:2007ry},
angular information is crucial to distinguish DM annihilation from
decay.  Whereas the rate of DM annihilation scales as the square of 
the DM density $\rho$, that of DM decay scales linearly with the
density, and consequently the angular distribution of the signal from
annihilation is expected to follow a steeper profile than that from
decay.  However the spatial distribution of DM substructure in a halo
also scales roughly as $\rho$.  Consequently, annihilation in this
component could produce a similar flattening in the angular
distribution of the observed emission as is expected for decay.  

Thus, in order to distinguish these possibilities, we
propose an observing strategy based on studying the angular variation
of the intensity and the energy spectrum of the signal.  From an
observational standpoint, a dramatic decrease in the observed
intensity between the center of the object and that at larger angles 
is a clear indicator of the simple case of annihilation in the smooth
halo only, while the observation of a shallow emission profile at all
angles would strongly suggest decay only.  On the other hand, the
observation of a bright central region but with the intensity falling
off more slowly in the outer regions is less straightforward to
interpret, as it could indicate annihilation with an important
contribution from substructure, or both annihilation and decay
contributing significantly.  In this case, we demonstrate that an
analysis of the energy spectrum of the signal as a function of angular
distance from the center of the object could provide the necessary
information to distinguish these possibilities. 

If only one process (annihilation or decay) produces a detectable
signal, the energy spectrum of the DM signal is the same from all
regions of the object, with the intensity varying according to how 
the rate of that process depends on the DM distribution.  If both
processes produce detectable signals, the energy spectrum of the total
signal varies according to the contribution from each process.  With
generality, we can assume that in this two-process scenario the
annihilation signal is always dominant in the inner regions of the 
object, with decay becoming more important at larger angles from the
center of the object.  Thus, we identify that both annihilation and
decay are present by observing a change in the energy spectrum of the
signal as a function of angle.  By measuring this change, the presence
of both annihilation and decay is confirmed, so by examination of the
signal in the inner and outer regions of the object, the degeneracy in
the DM particle mass could be broken.  In this case the DM lifetime
and annihilation cross section could also be determined from the
indirect measurement, up to uncertainties in the density profile and,
for the signal from the outer regions, uncertainties in the properties
of substructure.

In the following we illustrate the main points just outlined for the
case of gamma-ray observations of dwarf galaxies.  Dwarf galaxies are
extremely DM--dominated, with mass-to-light ratios in the range $100
\, M_\odot / L_\odot < M/L < 1000 \, M_\odot / L_\odot$~\cite{dGML}.
High DM densities coupled with minimal foregrounds due to a scarcity
of astrophysical gamma-ray sources make these objects excellent
targets for indirect DM searches in gamma-rays.  In addition, the
predicted emission from DM decay or annihilation in Milky Way dwarfs
has a relatively large angular extent ($\sim$ few degrees), which in
principle, makes it possible to map the angular distribution of an
observed signal.

We illustrate the proposed technique for three Milky Way dwarf
galaxies: Draco, Ursa Minor, and Sagittarius, which are among the most
optimistic for detection in gamma-rays (e.g.,
Refs.~\cite{Essig:2009jx, Abdo:2010ex}), and are all accessible
targets with current experiments.  We treat separately the
contributions from the smooth halo and substructure components to the
gamma-ray signal.  The smooth halo case  alone provides a lower limit
on the gamma-ray signal from annihilation for our assumed density
profile and represents the steepest angular emission profile.  On the
other hand, simulations indicate that a scaled-down host subhalo
population represents the maximum expected abundance of
sub-substructure~\cite{Diemand:2008in, Springel:2008cc}, so we model 
the subhalo population of each dwarf in this way to consider the upper
limits on the total annihilation flux and on the shallowness of the
angular emission profile in the annihilation case.  Let us note
however, that the properties of substructure in dwarf galaxy halos are
quite uncertain, but for completeness we also consider this potential
contribution.

We describe the mass distribution of the smooth DM halo of each dwarf
galaxy by a Navarro, Frenk and White (NFW) density profile~\cite{NFW}
and the collective emission from subhalos within the dwarf galaxy
halo, by summing over the contribution to the gamma-ray signal from
subhalos of all masses.  We assume that the density profile of each
subhalo can also be described by a NFW profile.  We refer the reader
to Ref.~\cite{PalomaresRuiz:2010pn} for details on the modeling of
the DM distribution and the measured and derived properties of the
selected dwarf galaxies.

\begin{figure*}[t] 
   \centering
   \includegraphics[width=.9\textwidth]{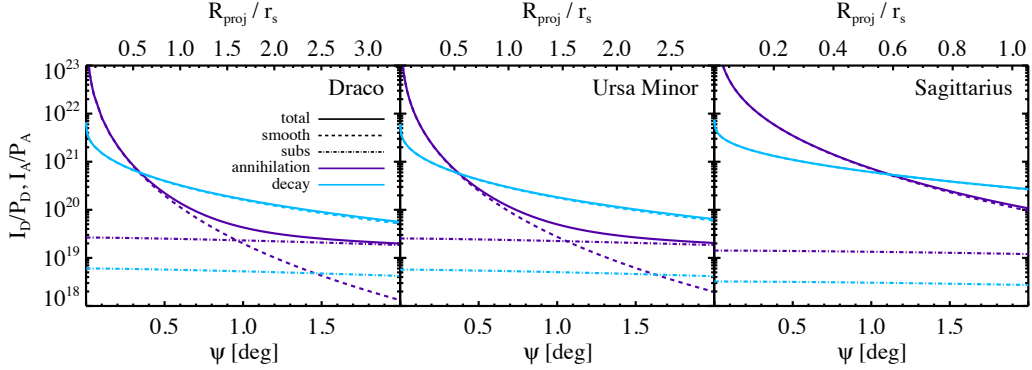} 
   \caption{\footnotesize Dependence of the intensity from DM decay
     ({\it blue}) and DM annihilation ({\it purple}) on line-of-sight
     direction $\psi$ from the center of the object for selected dwarf
     galaxies.  The contributions from the smooth halo ({\it dashed}),
     substructure ({\it dot-dashed}), and the total ({\it solid}) are 
     shown.  The corresponding projected radius $R_{\rm proj}$ in
     units of the halo scale radius $r_{\rm s}$ is labeled on the top
     axis.  For generality, the amplitudes of the curves for decay
     and annihilation have been scaled by the factors $P_{\rm D}$ and 
     $P_{\rm A}$ (defined in Ref.~\cite{PalomaresRuiz:2010pn})
     respectively, which depend on the assumed particle properties.
     $I_{\rm D}/P_{\rm D}$ is shown in units of GeV cm$^{-2}$
     sr$^{-1}$, and $I_{\rm A}/P_{\rm A}$ is shown in units of
     GeV$^{2}$ cm$^{-5}$ sr$^{-1}$.  From
     Ref.~\cite{PalomaresRuiz:2010pn}.
\label{fig:iprofiles}}
\end{figure*}

The angular dependence of the gamma-ray intensity from DM annihilation
and decay is shown in Fig.~\ref{fig:iprofiles} for our three example
dwarf galaxies.  The contributions from substructure and the smooth
halo are shown separately, along with the total of these signals from
each process.  The contribution from DM annihilation or decay in
substructure (blue and purple dot-dashed curves) tends to be nearly
parallel to the smooth halo contribution in the case of decay (blue
dashed curves) at angles $\gtrsim 1^{\circ}$.  Note that decay in
substructure is always subdominant relative to decay in the smooth
halo, even in the maximal substructure scenario we consider here.

\section{Results}

The first requirement in order to use this strategy is that the source 
is resolved as an extended source.  In particular, we assume that the
signal can be binned into several annuli centered on the source.  This
is in principle possible with the angular resolution of current
experiments ($\sim 0.1^{\circ}$ at the relevant energies) for
observations of dwarf spheroidal galaxies, since the angular extent of 
the predicted DM signal is as large as $\sim$ few degrees.  In
addition, this strategy requires that the signal in each annulus is
detected with sufficient statistics to reconstruct the energy
spectrum.  In the following we proceed under the assumption that these
conditions are met. 

\begin{figure*}[t] 
   \centering
   \includegraphics[width=0.6\textwidth]{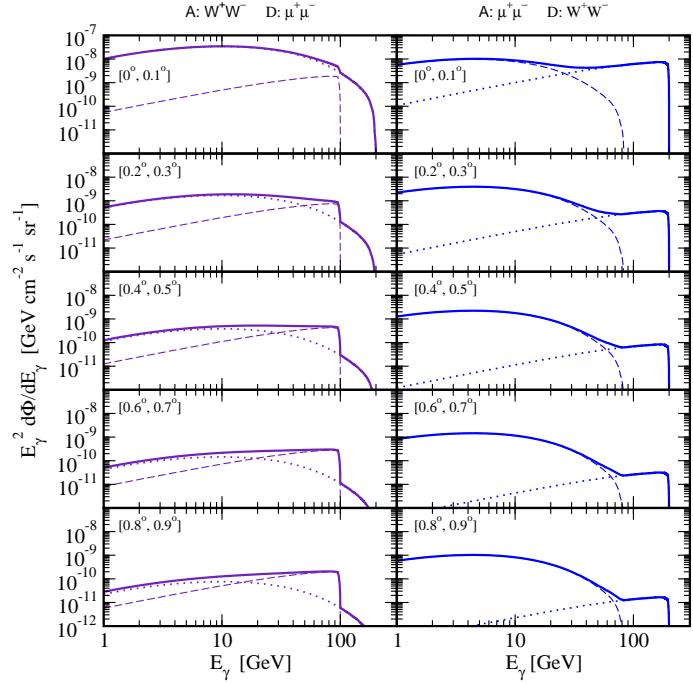}
   \caption{\footnotesize Energy spectra in different annuli centered
     on Draco for a DM mass of 200~GeV and for two combinations of
     channels. See text for details.  From
     Ref.~\cite{PalomaresRuiz:2010pn}.
\label{fig:annuli}} 
\end{figure*}

In Fig.~\ref{fig:annuli} we illustrate the proposed method for a
scenario in which both annihilation and decay contribute appreciably
to the observed signal from the Draco dwarf galaxy by showing the
energy spectrum as a function of the angle from the center of the
object.  The energy spectrum in alternating annuli of $0.1^{\circ}$
width centered on Draco is shown out to an angular radius of
$0.9^{\circ}$ (from top to bottom) for a DM particle mass of
$m_\chi=$~200~GeV.  Two combinations of channels are shown.  The left 
(right) column shows the case of annihilation into a soft (hard)
channel and decay into a hard (soft) one.  The channels $\mu^+ \mu^-$
and $W^+ W^-$ have been chosen as representative of hard and soft
channels, respectively.  In each panel, dashed lines represent the
contribution from decay, dotted lines represent that from annihilation,
and the thick solid lines represent the total contribution.  We have
taken $\langle \sigma v \rangle = 3 \times 10^{-26}$~cm$^3$~s$^{-1}$
and $\tau_\chi = 10^{29}$~s.  Note that although we have included the
contribution of substructure, it is a subdominant effect for both
annihilation and decay for the annuli considered in this figure (see
Fig.~\ref{fig:iprofiles}).  As expected, a significant change in the
spectrum is clearly seen in Fig.~\ref{fig:annuli} for both
combinations of channels at around $E = m_{\chi}/2$, i.e., the maximum
energy for photons from DM decay.  The spectral change is a signature
of both annihilation and decay contributing significantly to the signal.

\begin{figure*}[t] 
   \centering
   \includegraphics[width=0.45\textwidth]{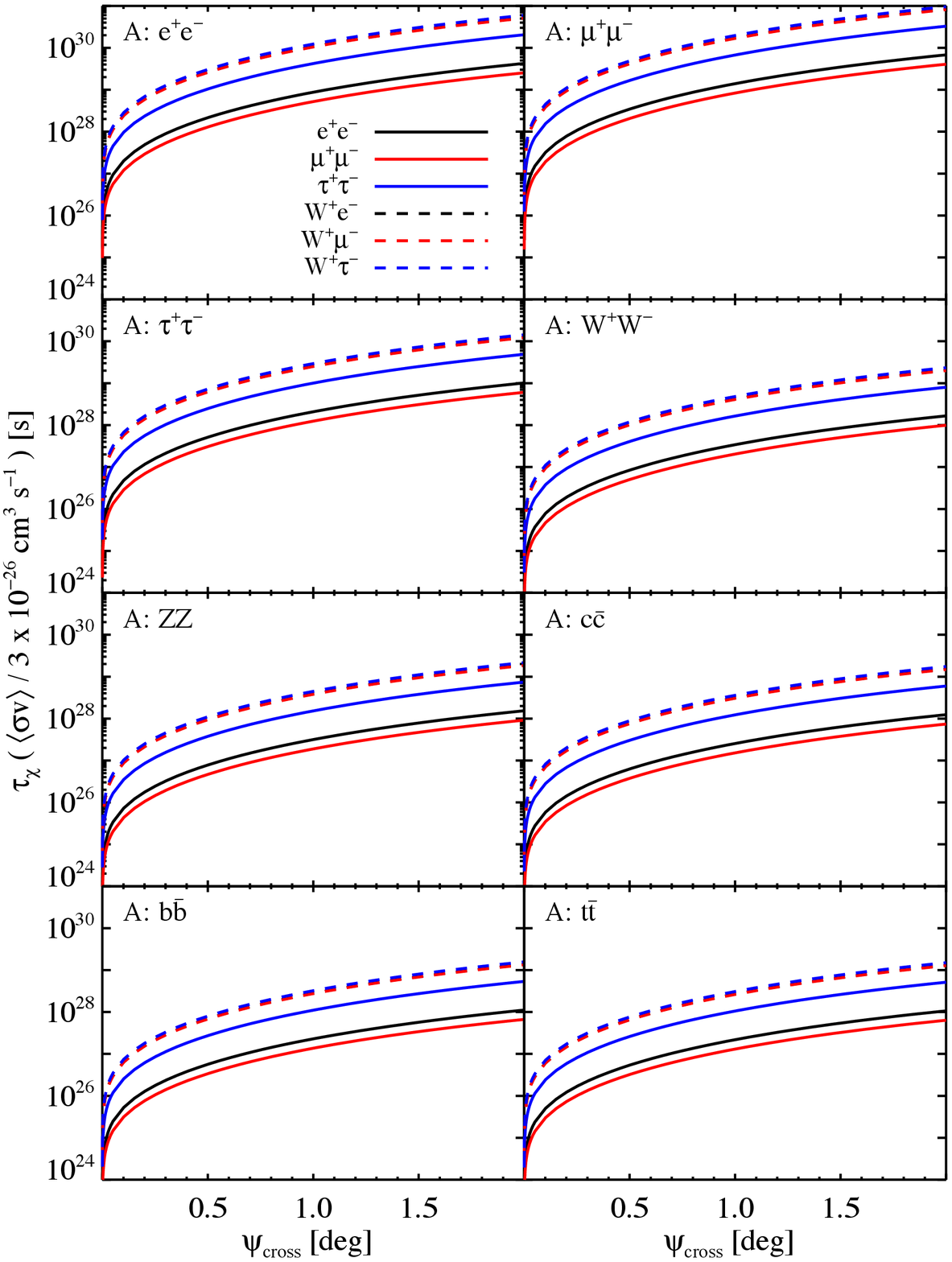}  
\includegraphics[width=0.45\textwidth]{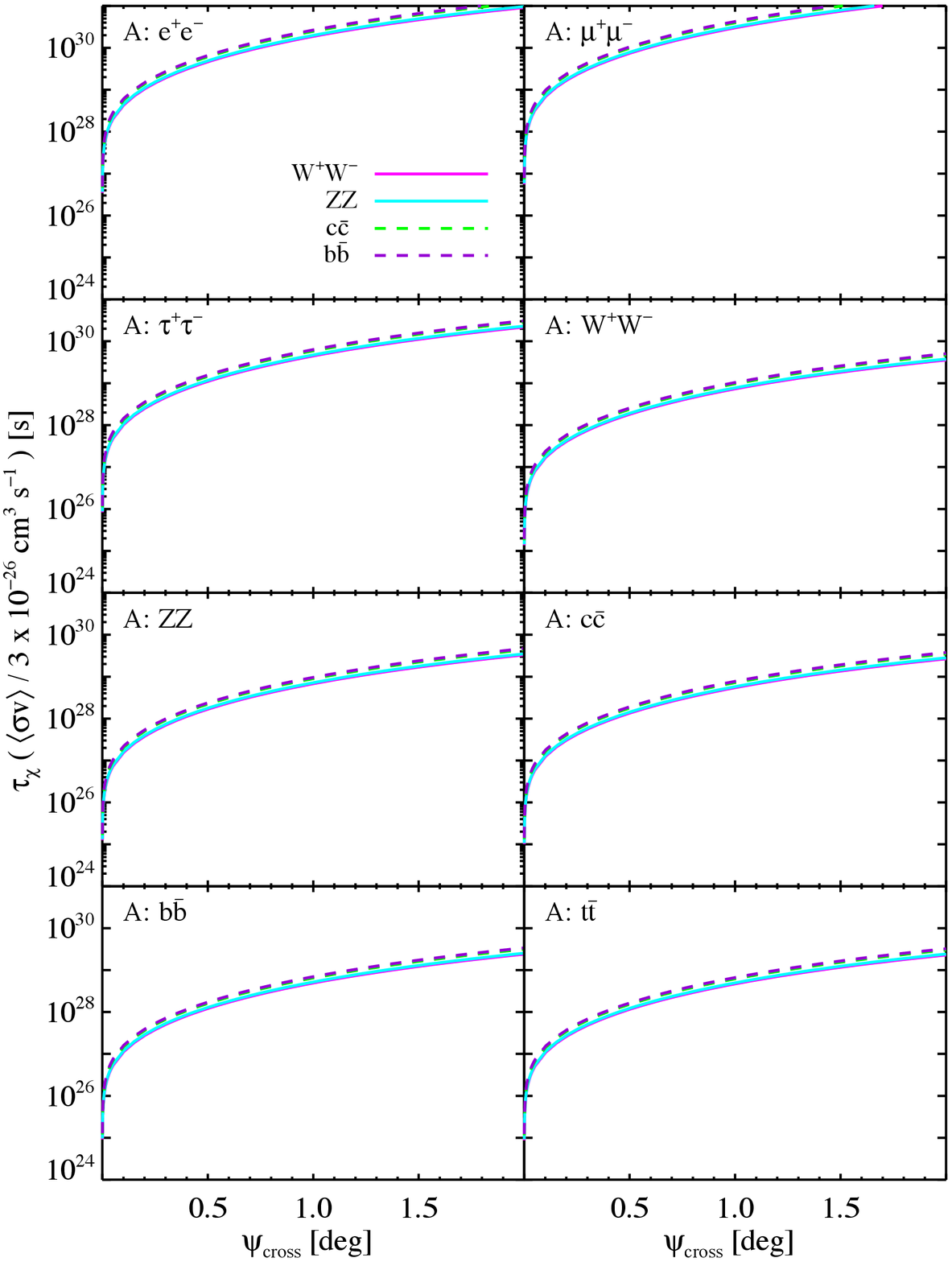} 
   \caption{\footnotesize Lifetime $\tau_\chi$ for $m_\chi =$~200~GeV
     at which the intensities from annihilation and decay for $E >
     1$~GeV are equal at an observation angle $\psi_{\rm cross}$ from
     the center of the dwarf galaxy, for Draco, without substructure.
     Each panel shows curves for a single annihilation channel,
     assuming decay into different channels (as labeled).  From
     Ref.~\cite{PalomaresRuiz:2010pn}.
\label{fig:example}} 
\end{figure*}

Fig.~\ref{fig:example} indicates the range of DM parameters which
would induce a transition between  annihilation and decay in the
angular range of $0^{\circ}$-$2^{\circ}$ in Draco, (similar results
are obtained for the other two dwarf galaxies).  Here we neglect the
contribution from substructure.  The curves indicate the value of the
DM lifetime at which the intensities from DM annihilation and decay
integrated above 1~GeV are equal at an observation angle $\psi_{\rm
  cross}$.  The results for DM decay into leptonic and semileptonic 
(hadronic and gauge boson) channels are shown in the left (right)
panels.  The annihilation channel for each panel is labeled.  In these
figures we assume $m_\chi = $~200~GeV and $\langle \sigma v \rangle
= 3 \times 10^{-26}$~cm$^3$~s$^{-1}$.  A larger cross section would
displace the curves downwards.  For a given $\psi_{\rm cross}$, above
the curves annihilation dominates and the emission profile is steeper,
while below the curves the dominant contributor is decay and the
profile is shallower.

The normalization of the curves depends on the relative photon yields
from annihilation and decay: for a given lifetime, the
annihilation-to-decay transition occurs further from the center of the
dwarf galaxy for channel combinations in which the ratio of the photon
yields from annihilation to decay is larger.  In each panel,
corresponding to a single annihilation channel, the variation in the
amplitude of the curves reflects the different photon yields for the
decay channels shown.  Decay via any of the hadronic or gauge boson
channels produces almost identical curves since the photon yields
above 1~GeV from these channels are similar, and these curves have the
highest normalization of any of the channels since their photon yields
are the highest. Similarly, there is little difference between the
curves for decay into any of the three semi-leptonic channels, and
these curves fall below the hadronic and gauge boson decay channel
curves.  The curves for decay into the leptonic channels show more
variation due to the larger variation in photon yields for these
channels, and as expected, fall below those for semi-leptonic and
hadronic and gauge boson channels due to their relatively low photon
yields.

For this energy threshold and an assumed cross section of $\langle
\sigma v \rangle = 3 \times 10^{-26}$~cm$^3$~s$^{-1}$, in order for
the transition to occur at an angle between $\sim 0.1^{\circ}$ and
$\sim 2^{\circ}$, the DM lifetime must be between $\sim 10^{25}$~s and
$10^{31}$~s, depending on the combination of channels.  For larger
values of the annihilation cross section, correspondingly smaller
values of the lifetime are needed.

\section{Conclusions}

In this talk we have outlined a strategy to constrain DM properties in
the event of the clear detection of an indirect signal from gamma-ray
observations of dwarf galaxies.  We addressed the question of how
scenarios of DM annihilation, decay, or both could be
distinguished, and what information could be obtained about the
intrinsic properties of the DM particle and its small-scale
distribution from this type of indirect measurement.  In summary, we
have shown that a DM particle with an annihilation cross-section and
lifetime just beyond the limits currently established could produce a
clear spectral change on an angular scale within the reach of future
experiments.  Ongoing observations by current and future experiments
will continue to improve the prospects for detecting and mapping a DM
signal in the coming years.

\end{document}